\documentclass[prd,twocolumn,floatfix,amsmath,nofootinbib,amssymb,floatfix]{revtex4}
\usepackage{graphicx,color,dcolumn,booktabs,bm}
\usepackage{longtable,lscape}
\usepackage{txfonts}
\usepackage{overpic}
\usepackage{amssymb}
\usepackage{indentfirst}
\usepackage{feynmf}   
\usepackage{slashed}  
\usepackage{cases}
\usepackage{color}
\usepackage{multirow}
\usepackage{epstopdf}
\usepackage{enumerate}
\usepackage{graphicx,color,dcolumn,booktabs,bm}
\usepackage[colorlinks, citecolor=blue,anchorcolor=red,menucolor=red, linkcolor=red,filecolor=red,urlcolor=blue,frenchlinks=red]{hyperref}

\graphicspath{{Figures/}} %

\begin{document}

\title{Confirming the existence of a new higher charmonium $\psi(4500)$ by the newly released data of $e^+e^- \to   K^+K^- J/\psi$}   
\author{Jun-Zhang Wang$^{1,2}$}\email{wangjzh2022@pku.edu.cn}
\author{Xiang Liu$^{2,3,4}$}\email{xiangliu@lzu.edu.cn}
\affiliation{$^1$School of Physics and Center of High Energy Physics, Peking University, Beijing 100871, China\\
$^2$School of Physical Science and Technology, Lanzhou University, Lanzhou 730000, China\\
$^3$Research Center for Hadron and CSR Physics, Lanzhou University $\&$ Institute of Modern Physics of CAS, Lanzhou 730000, China\\
$^4$Lanzhou Center for Theoretical Physics, Key Laboratory of Theoretical Physics of Gansu Province, and Frontiers Science Center for Rare Isotopes, Lanzhou University, Lanzhou 730000, China}

\date{\today}

\begin{abstract} 
Recently, the BESIII Collaboration published the Born cross section of $e^+e^- \to   K^+K^- J/\psi$, in which a new vector enhancement structure $Y(4500)$ was observed for the first time.  The mass of  the $Y(4500)$ resonance structure is in good agreement with our previous prediction for a $5S$-$4D$ mixing charmonium candidate $\psi(4500)$, while the measured width is $2\sigma$ larger than the theoretical estimate. In this work, we reanalyzed the cross section of $e^+e^- \to   K^+K^- J/\psi$ by introducing two theoretically predicted charmonium states $\psi(4220)$ and $\psi(4500)$ as intermediate resonances, and found that this width deviation problem can be resolved. Meanwhile, we found that the inclusion  of  two established charmonium states $\psi(4160)$ and $\psi(4415)$ can further improve the fit quality, with four sets of widely different solutions to $\Gamma_{e^+e^-} \mathcal{B}(\psi \to K^+K^- J/\psi)$. In order to better identify the properties of the $Y(4500)$, we calculated the charmonium hadronic transition to  final states $K^+K^- J/\psi$ within the charmed meson loop mechanism, where the fitted branching ratio of $\psi(4500) \to K^+K^- J/\psi$ can be well explained by the theoretical calculation, further supporting the new enhancement structure around 4.5 GeV as a higher charmonium $\psi(4500)$ that we predicted.   Finally, we also discussed the possible signal of another predicted charmonium candidate $\psi(4380)$ in $e^+e^- \to   K^+K^- J/\psi$. These research results should be an important step in constructing the charmonium family and thoroughly solving the charmoniumlike $Y$ problem.
\end{abstract}

\maketitle

\section{Introduction}

The discovery of a series of charmoniumlike $XYZ$ structures above the open-charm threshold has opened a new  perspective for the study of hadron spectroscopy in the last two decades.  Their unexpected masses and decay behaviors not only inspire extensive discussions on various exotic hadronic configurations, but also greatly challenge our inherent knowledge for the conventional charmonium spectroscopy (see review articles \cite{Chen:2016qju,Liu:2019zoy,Chen:2022asf,Guo:2017jvc,Olsen:2017bmm,Brambilla:2019esw} for comprehensive experimental and theoretical progresses). Therefore, these charmoniumlike states provide very a unique platform to help us to deepen our understanding of the non-perturbative behavior of the strong interaction.


As the important members of the $XYZ$ family, the observations of the abundant $Y$ states with $J^{PC}=1^{--}$ from direct $e^+e^-$ annihilation processes had aroused wide interest among experimentalists and theorists \cite{Chen:2016qju,Liu:2019zoy,Chen:2022asf}.  In particular,  the $Y(4260)$ was first reported in the cross section measurement of $e^+e^-\to \pi^+\pi^- J/\psi \gamma_{\mathrm{ISR}}$ by the BaBar experiment \cite{BaBar:2005hhc}. In 2017, the BESIII Collaboration showed, within more precise experimental data,  that the original peak signal referred to the $Y(4260)$ in $e^+e^-\to \pi^+\pi^-J/\psi$ can be replaced by two resonance structures, i.e., the $Y(4220)$ and $Y(4330)$ \cite{BESIII:2016bnd}.  Subsequently, the $Y(4220)$  was confirmed in other hidden-charm and open-charm decay channels, including $ \pi^+\pi^- h_c$ \cite{BESIII:2016adj}, $\omega \chi_{c0}$ \cite{BESIII:2019gjc}, $\pi^+\pi^-\psi(3686)$ \cite{BESIII:2017tqk}, $\eta J/\psi $ \cite{BESIII:2020bgb}, $\pi^0\pi^0 J/\psi$ \cite{BESIII:2020oph}, and $D^0D^{*-}\pi^{+}$ \cite{BESIII:2018iea}.  In the current version of the Particle Data Group (PDG), the average mass and width of the $Y(4220)$ are $m=4222.7\pm2.6$ MeV and $\Gamma=49\pm8$ MeV \cite{ParticleDataGroup:2022pth}.  In addition to the superstar $Y(4220)$, two higher states $Y(4360)$ and $Y(4660)$ were discovered in the invariant mass spectrum of $\psi(2S)\pi^+\pi^-\gamma_{\mathrm{ISR}}$ by Belle \cite{Belle:2007umv} and two other charmoniumlike structures $Y(4390)$ and $Y(4630)$ were reported in the $e^+e^- \to  \pi^+\pi^- h_c$ \cite{BESIII:2016adj} and $e^+e^- \to \Lambda_c \bar{\Lambda}_c$ \cite{Belle:2008xmh}, respectively.  
Despite many theoretical efforts to interpret the nature of these $Y$ states  \cite{Chen:2016qju,Liu:2019zoy,Chen:2022asf},  it is not an easy task to naturally assign so many vector $Y$ states in a unified theoretical framework, which should be very challenging.

Faced with the confused $Y$ problem,  the Lanzhou group has provided a complete solution scheme \cite{He:2014xna,Chen:2014sra,Chen:2017uof,Wang:2019mhs,Wang:2020prx,Qian:2021gby}. In 2017, a Fano-like interference picture was proposed to reproduce the line shape of  the $Y(4330)$ and $Y(4390)$ in hidden-charm channels, which were identified as non-resonant signals \cite{Chen:2017uof}. In Ref. \cite{Wang:2019mhs}, combined with the updated data of $Y$ states,  the Lanzhou group  found that the $Y(4220)$ can be treated as a good  scaling point to construct the $J/\psi$ family in an unquenched quark potential model, where the unquenched effect of the color confinement potential plays a crucial role in describing the higher charmonium spectrum above 4 GeV \cite{Wang:2019mhs}. In this framework, the $Y(4220)$ and $\psi(4415)$ can be assigned very well to the $J/\psi$ family under a $4S$-$3D$ and  $5S$-$4D$ mixing scheme, respectively. More importantly, two corresponding partner particles $\psi(4380)$ and $\psi(4500)$ were predicted. Furthermore, we showed some possible clues of the predicted $\psi(4380)$ in $e^+e^- \to  \pi^+\pi^- \psi(3686)$, which can just be related to the peak of the $Y(4360)$ state  \cite{Wang:2019mhs}. Therefore, our previous research results gave a unified picture to unify all reported $Y$ states below 4.5 GeV and predicted the existence of a new charmonium state $\psi(4500)$, which missed in experiments.   


Recently, the BESIII Collaboration published the cross section measurement of $e^+e^- \to   K^+K^- J/\psi$ corresponding to $E_{cm}=4.127-4.600$ GeV, where two resonant structures were observed \cite{BESIII:2022joj}. The resonance parameters of the first structure are close to those of the $Y(4220)$, while the mass and width of the second structure are fitted with a statistical significance greater than $8\sigma$ to be $4484.7 \pm13.3 \pm 24.1$ MeV and $111.1\pm30.1\pm5.2$ MeV, respectively. Accordingly, this new charmoniumlike structure was named the $Y(4500)$ by BESIII \cite{BESIII:2022joj}. To our surprise,  the reported mass position of $Y(4500)$ is just consistent with the theoretical mass range of 4489-4529 MeV for charmonium $\psi(4500)$ predicted by the Lanzhou group  \cite{Wang:2019mhs}.  Thus, it is natural to speculate that this new experimental discovery is closely  linked to the missing charmonium state $\psi(4500)$. 

In order to further identify the nature of the $Y(4500)$, in this work, we have performed a careful analysis for the cross section from $e^+e^- \to   K^+K^- J/\psi$ by introducing two theoretically predicted charmonium states $\psi(4220)$ and $\psi(4500)$ as intermediate resonances, where the divergence between the large experimental width of the $Y(4500)$ and the corresponding small theoretical width can be well clarified. Furthermore, we find out a four-resonance scheme to improve the overall fitting quality, where  the contributions of two well-established charmonium states $\psi(4160)$ and $\psi(4415)$ have been considered.  In this improved fitting scenario,  four sets of very different solutions to the product of the di-lepton width $\Gamma(\psi \to e^+e^-)$ and the branching ratio $\mathcal{B}(\psi \to K^+K^- J/\psi)$ can be obtained, providing valuable information for testing higher charmonium assignments above 4.2 GeV. Inspired by these fitted solutions associated with the $K^+K^- J/\psi$ channel, we studied the hidden charm decays of  $\psi \to K^+K^- J/\psi$ within the charmed meson loop mechanism, where $\psi=\psi(4160)$, $\psi(4220)$, $\psi(4380)$, $\psi(4415)$, and $\psi(4500)$. We found that the fitted branching ratio of $\psi(4500) \to K^+K^- J/\psi$ can be reproduced well in our scenario, given a reasonable range of parameters. Meanwhile, after further making comparison of the fitted solutions with $\Gamma(\psi \to e^+e^-) \mathcal{B}(\psi \to K^+K^- J/\psi)$ and the corresponding predictions for other $\psi$ states included in the fitting scenario, we find only one suitable solution that can be supported by the theoretical calculation. Thus, these theoretical results further confirm the presence of a $5S$-$4D$ mixing charmonium $\psi(4500)$ and support our proposal to solve the  charmoniumlike $Y$ problem.
Finally, based on the theoretical prediction of the branching ratio $\mathcal{B}(\psi(4380) \to K^+K^- J/\psi)$, we also discussed the possible signal of another predicted charmonium $\psi(4380)$ in the $e^+e^- \to   K^+K^- J/\psi$ process.

This paper is organized as follows. In the following section, we focus on the analysis on the cross section of $e^+e^- \to   K^+K^- J/\psi$ (see Sec. \ref{sec2}). Based on the fitting results of the cross section, in Sec. \ref{sec3}, we calculate the branching ratios of the involved intermediate charmonium decays to the hidden charm channel $K^+K^- J/\psi$ within the charmed meson loop mechanism, and present the related discussions. Finally, this work ends with a short summary in Sec. \ref{sec4}.

\section{The analysis to the cross section data from $e^+e^- \to   K^+K^- J/\psi$}\label{sec2}

As mentioned in the Introduction, the BESIII Collaboration observed two charmoniumlike structures $Y(4220)$ and $Y(4500)$ in the cross section distribution of $e^+e^- \to   K^+K^- J/\psi$ against the center-of-mass energy. It is worth noting that the experimental width of the $Y(4220)$ in the $K^+K^- J/\psi$ channel is $72.9\pm6.1\pm30.8$ MeV \cite{BESIII:2022joj}, which is larger than the average width ($49\pm8$ MeV) of the $Y(4220)$ given in the Particle Data Group (PDG) \cite{ParticleDataGroup:2022pth} and our theoretical width ($23-30$ MeV) for a $4S$-$3D$ mixing charmonium $\psi(4220)$ \cite{Wang:2019mhs}. In addition, BESIII \cite{BESIII:2022joj} also pointed out that the measured width $\Gamma=111.1\pm30.1\pm5.2$ MeV of the $Y(4500)$ is $2\sigma$ larger than our predicted width of $36\sim45$ MeV for a higher $5S$-$4D$ mixing charmonium $\psi(4500)$ \cite{Wang:2019mhs}. Motivated by this width discrepancy problem, in this section, we reanalyze the cross section of the $e^+e^- \to   K^+K^- J/\psi$ reaction by considering the resonance contributions from the theoretically established higher charmonium states.

\begin{figure}[t]
	\includegraphics[width=8.0cm,keepaspectratio]{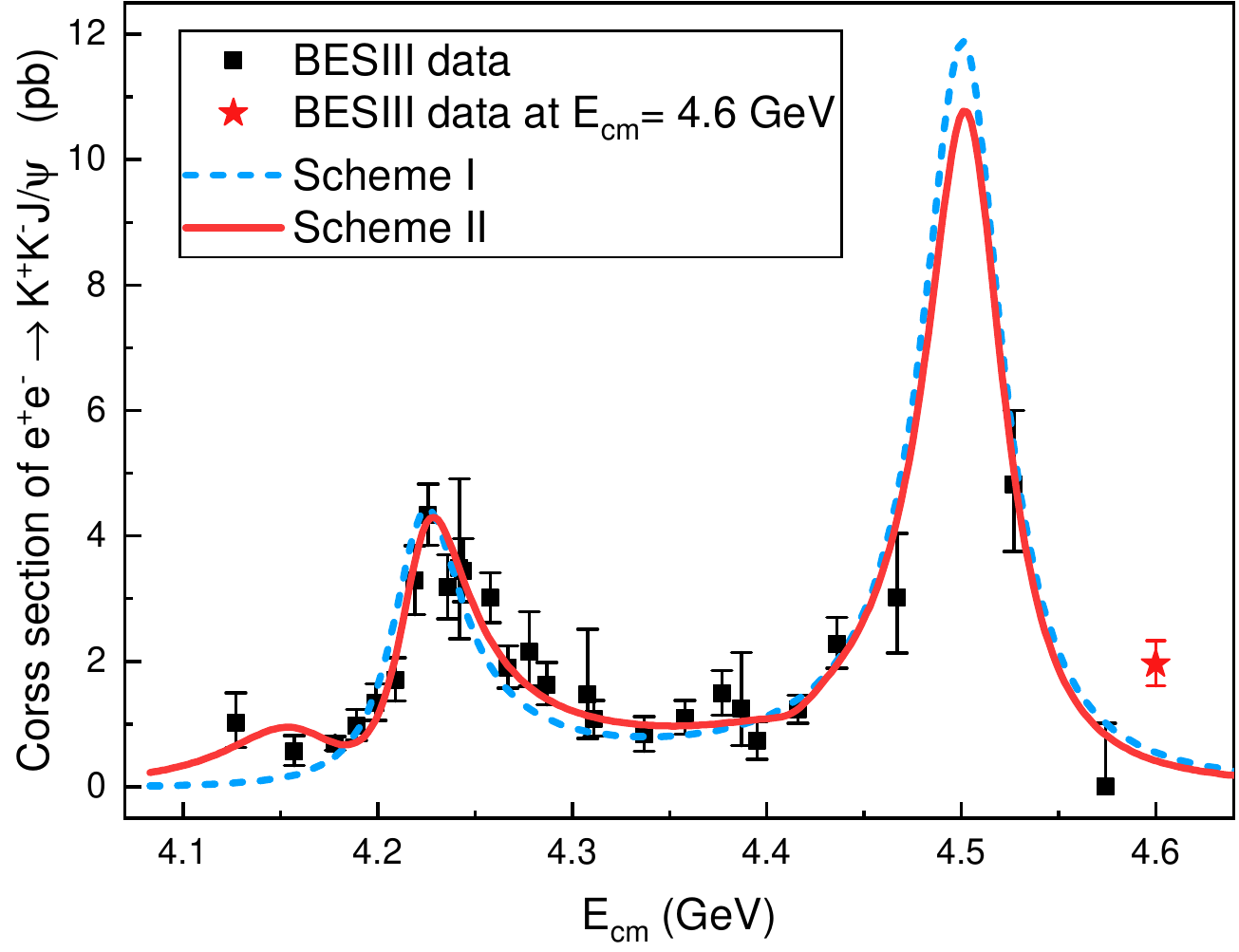}
	\caption{Our fitting result to the cross section of the $e^+e^- \to   K^+K^- J/\psi$ process between $E_{cm}=$4.127 to 4.600 GeV from BESIII. Here, the red five-point star represents the BESIII data point at 4.6 GeV  \cite{BESIII:2022joj}.   }\label{Fit}
\end{figure}

\begin{table}[h]
  \begin{ruledtabular}
  \caption{ The input of the masses and widths for these theoretically established vector charmonium states between 4.1 and 4.6 GeV, which are taken from Ref. \cite{Wang:2019mhs}.}\label{masswidth}
  \centering
  \begin{tabular}{cccc}
     States & Assignment\footnote{The $\psi$ states with single  prime and double prime stand for the lower and higher $S$-$D$ mixing charmonium states, respectively. } & Mass (MeV) & Width (MeV)  \\
    \hline
    $\psi(4160)$ & $\psi_{3S-2D}^{\prime\prime}$ & $4159\pm20$ \cite{DASP:1978dns} & $78\pm20$ \cite{DASP:1978dns}\\
    $\psi(4220)$ & $\psi_{4S-3D}^{\prime}$ & $4222\pm3.1\pm1.4$ \cite{BESIII:2016bnd}  & $44.1\pm4.3\pm2.0$ \cite{BESIII:2016bnd}  \\
     $\psi(4380)$ & $\psi_{4S-3D}^{\prime\prime}$ & $4364\sim4440$ \cite{Wang:2019mhs} & $80$\footnote{The typical values based on the estimate ranges from theoretical model \cite{Wang:2019mhs}. }$$ \\
     $\psi(4415)$ & $\psi_{5S-4D}^{\prime}$ & $4414\pm7$ \cite{Siegrist:1976br}& $33\pm10$ \cite{Siegrist:1976br}  \\
     $\psi(4500)$ & $\psi_{5S-4D}^{\prime\prime}$ & $4489\sim4529$ \cite{Wang:2019mhs}  & $50^b$  \\
  \end{tabular}
  \end{ruledtabular}
\end{table}

\begin{table*}[t]
  \begin{ruledtabular}
  \caption{ The fitted parameters in four-resonance fitting scheme II. The phase angles $\phi_1$, $\phi_2$ and $\phi_3$ are relevant for the $\psi(4500)$, $\psi(4415)$ and $\psi(4160)$, respectively. }\label{parameter}
  \centering
  \begin{tabular}{ccccc}
     Parameters & Solution $A$ & Solution $B$ & Solution $C$ & Solution $D$   \\
    \hline
    $\Gamma(\psi(4220) \to e^+e^-)\mathcal{B}(\psi(4220) \to K^+K^- J/\psi)$ ~(eV) & $0.557\pm0.045$ & $0.538\pm0.020$ & $0.193\pm0.024$ & $0.186\pm0.021$\\
    $\Gamma(\psi(4500) \to e^+e^-)\mathcal{B}(\psi(4500) \to K^+K^- J/\psi)$ ~(eV) & $0.915\pm0.117$ & $0.774\pm0.082$ & $0.895\pm0.135$ & $0.757\pm0.091$ \\
    $\Gamma(\psi(4415) \to e^+e^-)\mathcal{B}(\psi(4415) \to K^+K^- J/\psi)$ ~(eV) & $0.259\pm0.026$ & $(6.440\pm13.419)\times10^{-4}$ & $0.252\pm0.027$ & $(6.343\pm9.804)\times10^{-4}$ \\
    $\Gamma(\psi(4160) \to e^+e^-)\mathcal{B}(\psi(4160) \to K^+K^- J/\psi)$ ~(eV) & $0.306\pm0.047$ & $0.295\pm0.022$ & $0.0782\pm0.0247$ & $0.0756\pm0.0145$ \\
    $\phi_{1}$ ~(rad) &  $3.670\pm0.412$    & $4.132\pm0.148$ & $2.856\pm0.417$ & $3.318\pm0.407$\\
    $\phi_{2}$ ~(rad) &  $5.568\pm0.293$     & $6.012\pm1.077$ &  $4.810\pm0.304$ & $5.263\pm1.079$ \\
    $\phi_{3}$ ~(rad) &  $1.954\pm0.106$     & $1.993\pm0.036$ & $0.385\pm0.180$ & $0.424\pm0.130$ \\
    $m_{\psi(4500)}$ (MeV) & \multicolumn{4}{c}{$4504\pm4$} \\
  \end{tabular}
  \end{ruledtabular}
\end{table*}

In general, the cross section of $e^+e^- \to   K^+K^- J/\psi$ comes mainly from two types of production mechanisms. The first one is  the direct continuum term and usually provides a background contribution. According to the experimental data of $e^+e^- \to   K^+K^- J/\psi$ from BESIII, its continuum background contribution is very small \cite{BESIII:2022joj}, which is negligible in our following analysis. Therefore, the $e^+e^- \to   K^+K^- J/\psi$ reaction should be dominated by the second contribution, which is mediated by a series of intermediate charmoniumlike resonances.  The contribution of  a genuine intermediate charmonium  to  $e^+e^- \to   K^+K^- J/\psi$ can be described by a phase space corrected Breit-Wigner function, i.e., 
\begin{eqnarray}
\mathcal{M}_{\psi}(s)&=&\frac{\sqrt{12\pi \Gamma(\psi \to e^+e^-)\mathcal{B}(\psi \to K^+K^- J/\psi)\Gamma_{\psi}}}{s-m_{\psi}^2+im_{\psi}\Gamma_{\psi}} \nonumber \\
&&\times \sqrt{\frac{\Phi_{2\to3}(s)}{\Phi_{2\to3}(m_{\psi}^2)}},
\end{eqnarray}
where $m_{\psi}$ and $\Gamma_{\psi}$ are the mass and width of the intermediate charmonium, respectively, and $\Phi_{2\to3}(s)$ is the phase space. The total amplitude of $e^+e^- \to   K^+K^- J/\psi$ can be written as 
\begin{eqnarray}
\mathcal{M}_{\mathrm{Total}}(s)=\mathcal{M}_{\psi_0}(s)+\sum_ie^{i\phi_i}\mathcal{M}_{\psi_i}(s),
\end{eqnarray}
where $\phi_i$ denotes the phase angle between the resonance amplitude $\mathcal{M}_{\psi_i}(s)$ and $\mathcal{M}_{\psi_0}(s)$. And then, the total cross section can be represented by $\sigma(s)=|\mathcal{M}_{\mathrm{Total}}(s)|^2$.

With the above preparations,  we can make a concrete fit to the experimental data of the cross section from $e^+e^- \to   K^+K^- J/\psi$. Similar to the BESIII analysis, we first considered the contributions of two intermediate charmonia $\psi(4220)$ and $\psi(4500)$ in the fitting scheme I, whose masses and widths as input quantities are summarized in Table \ref{masswidth}. There is a large theoretical uncertainty for the calculated resonance parameters of the $\psi(4500)$ state \cite{Wang:2019mhs}. In order to reduce the fitted parameters as much as possible, we set the mass and width of the $\psi(4500)$  as a bounded parameter and a typical input value, respectively, as shown in Table \ref{masswidth}. The fitted line shape of the cross section from $e^+e^- \to   K^+K^- J/\psi$, drawn by a dashed-blue curve, is shown in Fig. \ref{Fit}. Interestingly, we found that the enhancement structure around 4.5 GeV in $e^+e^- \to   K^+K^- J/\psi$ can indeed be reproduced by a narrower $\psi(4500)$ state. The reason for this is that a low cross section data point at 4.574 GeV was overlooked and instead a jump point at 4.600 GeV was included in the BESIII analysis \cite{BESIII:2022joj}. As discussed in Refs. \cite{Wang:2020prx,Qian:2021gby}, there should be  abundant vector charmonium candidates around 4.6 GeV, which can only be related to the charmoniumlike structures $Y(4630)$ and $Y(4660)$. Thus, we suspect that their contributions are very likely to enhance the cross section at 4.600 GeV and then affect the measured width of the new structure around 4.5 GeV. At the same time, combined with the fact of a lower experimental point at 4.574 GeV \cite{BESIII:2022joj}, the width deviation problem of the $\psi(4500)$ should be well resolved, which is worth emphasizing here.

The main feature of the cross section from $e^+e^- \to   K^+K^- J/\psi$ can be described in principle in the fitting scheme I, but it should be mentioned that more refined description of the experimental data around 4.2 GeV is still lacking, which can also be reflected in the value of $\chi^2/d.o.f.=2.23$ in scheme I. From another point of view, it is difficult to explain why there are no obvious contributions from two well-established charmonium states $\psi(4160)$ and $\psi(4415)$ in $e^+e^- \to   K^+K^- J/\psi$. Based on these two motivations, we further performed a four-resonance fitting scheme II, taking into account the contributions of the $\psi(4160)$ and $\psi(4415)$. The best fitting results in scheme II, represented by a red curve, are shown in Fig. \ref{Fit}, where a significantly improved $\chi^2/d.o.f.=1.07$ can be obtained. A direct comparison in Fig. \ref{Fit} shows that the line shape can be reproduced very well in scheme II. Our analysis also revealed an unremarkable resonance signal of $\psi(4160)$ in $e^+e^- \to   K^+K^- J/\psi$, which should be focused on in future BESIII and Belle II experiments. Here, it should be noted  that a slightly smaller mass of the $\psi(4160)$ can give  a better fit to the present data. However, the smaller mass does not seem to be supported by other measurements \cite{ParticleDataGroup:2022pth}, which should be clarified by more experimental data around 4.15 GeV. In addition, we found that the interference effect plays an important role in broadening the distribution of the resonance signal associated with the $\psi(4220)$, which can explain the corresponding large width of $72.9\pm6.1\pm30.8$ MeV in the experimental fit \cite{BESIII:2022joj}. 

The relevant fitted parameters in scheme II are listed in Table \ref{parameter}, where the mass of the $\psi(4500)$ was determined to be $4504\pm4$ MeV, and four sets of widely different solutions to the product of the di-lepton width $\Gamma(\psi \to e^+e^-)$ and the branching ratio $\mathcal{B}(\psi \to K^+K^- J/\psi)$ can be obtained. In the following section, we focus on the hidden-charm decays of  $\psi \to K^+K^- J/\psi$ and investigate whether these fitted branching ratios can be understood, which is necessary for further confirmation of higher charmonium assignments in the $J/\psi$ family.

\section{ Hidden-charm decays of $\psi \to K^+K^- J/\psi$ }\label{sec3}

\subsection{Charmed meson loop mechanism}

\begin{figure}[th]
	\includegraphics[width=8.5cm,keepaspectratio]{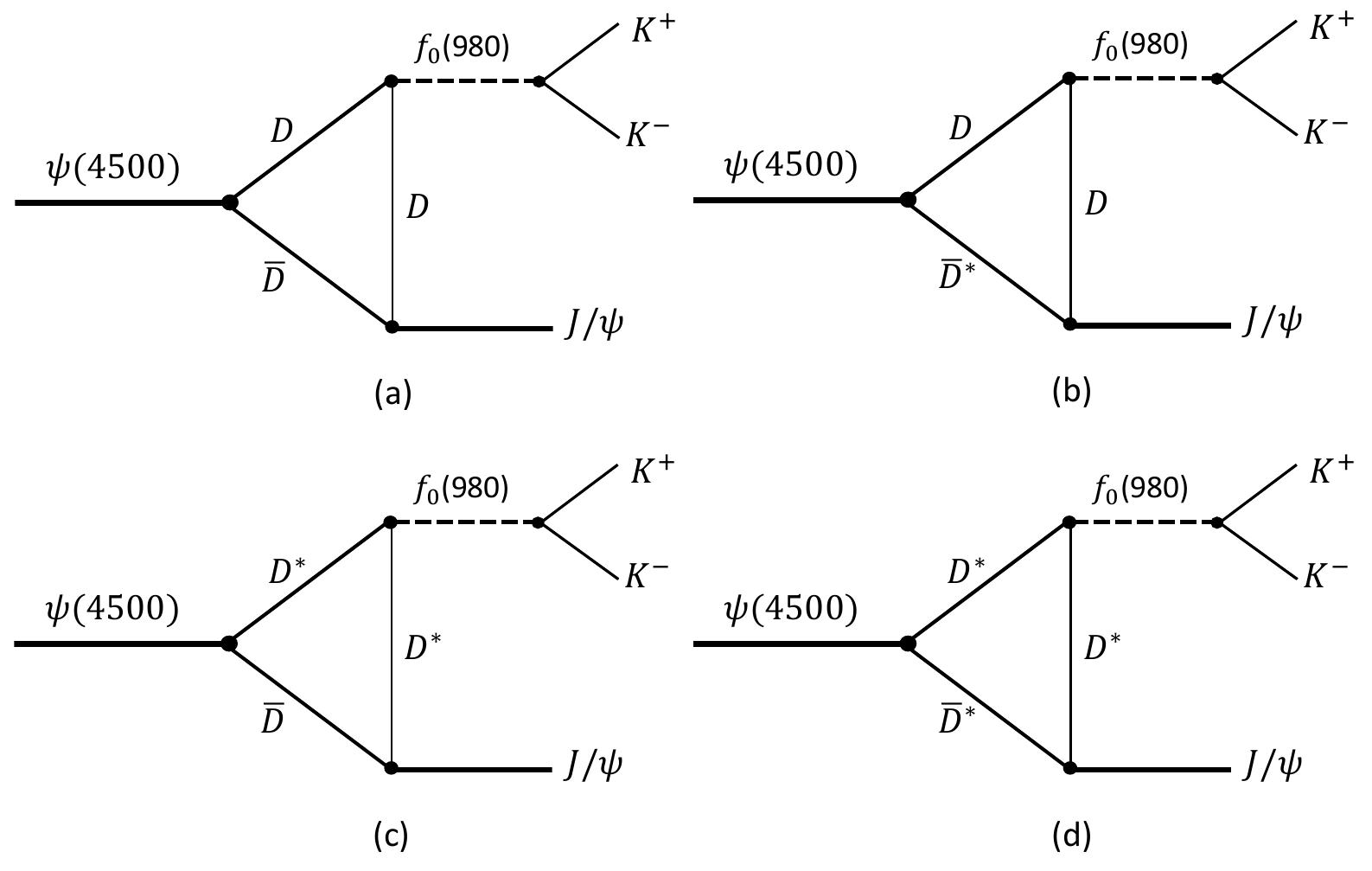}
	\caption{ The schematic charmed meson loop contributions to the hidden-charm decay  $\psi(4500) \to K^+K^- J/\psi$, which are the same for other higher charmonia $\psi(4160)$, $\psi(4220)$, $\psi(4380)$ and $\psi(4415)$.   }\label{triangle}
\end{figure}

The hidden-charm decay is usually an important class of  decay modes for charmonium. However, it is difficult to calculate in the first-principle framework of quantum chromodynamics (QCD), where the hidden-charm decay like $\psi \to K^+K^- J/\psi$ is involved in the non-perturbative problem of QCD. Here, an accessible approach to estimate the hidden-flavor decay ratio of heavy quarkonium is through the meson loop mechanism \cite{Lipkin:1986av,Liu:2006dq,Liu:2009dr,Zhang:2009kr,Guo:2009wr}.  For higher charmonium states above the threshold of a charmed meson pair, it should be noted that their absolutely dominant decay modes are the open-charm two-body decays to charmed meson pairs. Thus, it is quite natural that the produced charmed meson pairs can further transit into final states of a charmonium plus various light mesons by exchanging an intermediate charmed meson, the so-called charmed meson loop mechanism. The charmed meson loop contributions have been widely applied to explain various hidden-charm decay widths of the established charmonium states \cite{Liu:2009dr,Zhang:2009kr,Guo:2009wr,Chen:2012nva,Wang:2015xsa}.  The  Feynman diagrams of  $\psi \to K^+K^- J/\psi$ via the charmed  meson loop are shown in Fig. \ref{triangle}. To evaluate these diagrams, we have adopted the effective Lagrangian approach.
 In the following, we take $\psi=\psi(4500)$ as an example to illustrate the theoretical framework, which is also applicable to other higher $\psi$ states.

The first coupling vertex involved in the $\psi(4500)D^{(*)}\bar{D}^{(*)}$ interaction as shown in Fig. \ref{triangle} can be constructed based on the heavy quark symmetry \cite{Casalbuoni:1996pg,Colangelo:2003sa}.   For $S$-wave and $D$-wave charmonium multiplets, they can be expressed by \cite{Casalbuoni:1996pg,Colangelo:2003sa}
\begin{eqnarray}
\mathcal{J}_s=\frac{1+\slashed{v}}{2}\Big[\psi_s^\mu\gamma_\mu-\eta_c\gamma_5\Big]\frac{1-\slashed{v}}{2}
\end{eqnarray}
and
\begin{eqnarray}
\mathcal{J}^{\mu\lambda}_d&=&\frac{1+\slashed{v}}{2}\Bigg[\psi_3^{\mu\lambda\alpha}\gamma_\alpha+\frac{1}{\sqrt{6}}\Big(\epsilon^{\mu\alpha\beta\rho}v_\alpha\gamma_\beta \psi_{2\rho}^\lambda+\epsilon^{\lambda\alpha\beta\rho}v_\alpha\gamma_\beta \psi_{2\rho}^\mu\Big)\nonumber\\
&&+\frac{\sqrt{15}}{10}\Big[(\gamma^\mu-v^\mu)\psi_d^\lambda+(\gamma^\lambda-v^\lambda)\psi_d^\mu\Big]\nonumber\\
&&-\frac{1}{\sqrt{15}}\Big(g^{\mu\lambda}-v^\mu v^\lambda\Big)\gamma_\alpha\psi_d^\alpha+\eta_{c2}^{\mu\lambda}\gamma_5\Bigg]\frac{1-\slashed{v}}{2}, 
\end{eqnarray}
respectively, where $v^\mu$ is the 4-velocity of the heavy states. Based on the heavy quark symmetry, the general effective Lagrangians describing the interactions among the open-charm mesons and the $S$-wave or $D$-wave charmonium multiplet are
\begin{eqnarray} 
\mathcal{L}_s &=&g_1 \mathrm{Tr}\Big[\mathcal{J}_s\bar{H}_{2}\overleftrightarrow{\partial}_\mu\gamma^\mu\bar{H}_{1}\Big]+\mathrm{H.c.},\nonumber\\
\mathcal{L}_d &=&g_2 \mathrm{Tr}\Big[\mathcal{J}_d^{\mu\lambda}\bar{H}_{2}\overleftrightarrow{\partial}_\mu\gamma_\lambda\bar{H}_{1}\Big]+\mathrm{H.c.} , \label{generalL}
\end{eqnarray}
where $H_1$ and $H_2$ represent the current associated with the spin doublet of charmed meson field  \cite{Casalbuoni:1996pg,Colangelo:2003sa}. 
Since the $\psi(4500)$ is an $S$-$D$ wave mixing charmonium, which can be written as 
$\psi(4500)=-\sin\theta|5^3S_1\rangle+\cos\theta|4^3D_1\rangle$ \cite{Wang:2019mhs},
 so two types of effective Lagrangians are needed. Their concrete interaction expressions can be obtained by expanding the general Lagrangians in Eq. (\ref{generalL}), i.e., \cite{Casalbuoni:1996pg,Colangelo:2003sa,Wang:2016qmz} 
 \begin{eqnarray}\label{eq:lagrangians}
      \mathcal{L}_{\psi_s \mathcal{D}^{(*)}\mathcal{D}^{(*)}} 
       && = -ig_{\psi_s DD}\psi^\mu_s ( \mathcal{D}^{\dagger}\overleftrightarrow{\partial}_{\mu} \mathcal{D} ) \nonumber \\
       && \quad +g_{\psi_s DD^*}\epsilon^{\mu\nu\alpha\beta}\partial_\mu\psi_{s\nu} ( \mathcal{D}^{\dagger} \overleftrightarrow{\partial}_\alpha \mathcal{D}^*_{\beta}-\mathcal{D}_\beta^{*\dagger}\overleftrightarrow{\partial}_\alpha \mathcal{D} ) \nonumber\\
       && \quad +ig_{\psi_s D^*D^*}\psi^\mu_s ( \partial^\nu \mathcal{D}_\mu^{*\dagger} \mathcal{D}^*_\nu-\mathcal{D}_\nu^{*\dagger}\partial^\nu \mathcal{D}_\mu^*   
  \nonumber  \\&&\quad +\mathcal{D}^{*\nu\dagger}\overleftrightarrow{\partial}_{\mu} \mathcal{D}_\nu^* ),  
   \end{eqnarray}
   \begin{eqnarray}\label{eq:lagrangiand}
     \mathcal{L}_{\psi_d \mathcal{D}^{(*)}\mathcal{D}^{(*)}} 
       && = ig_{\psi_d DD}\psi^\mu_d ( \mathcal{D}^{\dagger}\overleftrightarrow{\partial}_{\mu} \mathcal{D} ) \nonumber\\
       && \quad -g_{\psi_d DD^*}\epsilon^{\mu\nu\alpha\beta}\partial_\mu\psi_{d\nu} ( \mathcal{D}^{\dagger} \overleftrightarrow{\partial}_\alpha \mathcal{D}^*_{\beta}-\mathcal{D}_\beta^{*\dagger}\overleftrightarrow{\partial}_\alpha \mathcal{D} ) \nonumber\\
       && \quad +ig_{\psi_d D^*D^*}\psi^\mu_d ( \partial^\nu \mathcal{D}_\mu^{*\dagger} \mathcal{D}^*_\nu-\mathcal{D}_\nu^{*\dagger}\partial^\nu \mathcal{D}_\mu^*   \nonumber \\
      &&\quad +4\mathcal{D}^{*\nu\dagger}\overleftrightarrow{\partial}_{\mu} \mathcal{D}_\nu^* ), 
 \end{eqnarray}
 where $\psi_s$ and $\psi_d$ stand for a $S$-wave and $D$-wave vector charmonium states, respectively, and $\mathcal{D}^{(*)}$ is charmed meson doublet $(D^{(*)0},D^{(*)+})^T$.  The effective Lagrangians related to scalar meson $f_0(980)$ can be written as \cite{Meng:2007cx,Meng:2007tk,Meng:2008dd}
 \begin{equation}\label{eq:lagrangianf0}
  \begin{split}
   \quad \mathcal{L}_{ \mathcal{D}^{(*)}\mathcal{D}^{(*)}f_0} 
       & = -g_{DDf_0} \mathcal{D}^{\dagger} \mathcal{D}f_0+g_{D^*D^*f_0} \mathcal{D}^{*\dagger}_{\mu} \mathcal{D}^{*\mu}f_0, \\
       \quad \mathcal{L}_{ f_0KK}&=g_{f_0KK} f_0K^{\dagger}K,
 \end{split}
 \end{equation}
 where $g_{DDf_0}=g_{D^*D^*f_0}=\sqrt{2}m_{D^*}g_{\pi}/\sqrt{6}$ with $g_{\pi}=3.73$ \cite{Meng:2007tk,Meng:2008dd}, and the $g_{f_0KK}=2.73$ GeV$^{-1}$ can be evaluated by the partial width of $f_0(980)\to K^+K^-$ \cite{ParticleDataGroup:2022pth}. For the interaction vertices of $J/\psi D^{(*)}\bar{D}^{(*)}$, it is obvious that their effective Lagrangians have the same coupling form as those in Eq. (\ref{eq:lagrangians}), where the coupling constants $g_{J/\psi DD}=7.44$, $g_{J/\psi DD^*}=3.84$ GeV$^{-1}$ and $g_{J/\psi D^*D^*}=8.00$ were obtained by the vector meson dominance \cite{Achasov:1994vh,Deandrea:2003pv}.

Based on the above effective Lagrangians, the decay amplitudes of $\psi(4500)(p) \to K^+(p_3)K^-(p_4) J/\psi(p_5)$ in Fig. \ref{triangle} can be given by 
\begin{eqnarray}
\mathcal{A}_{(a)}^{D\bar{D}}&=&-(i)^3\int\frac{d^4q}{(2\pi)^4}(\sin\theta g_{\psi_s DD}+\cos\theta g_{\psi_d DD})g_{J/\psi DD} \nonumber \\
&\times&(p_{5\alpha}-p_{3\alpha}-p_{4\alpha}-2q_{\alpha})\epsilon_{\psi(4500)}^{\alpha}(p_{5\nu}-2q_{\nu})\epsilon_{J/\psi}^{\nu*} \nonumber \\
&\times& \frac{1}{p_1^2-m_{D}^2}\frac{1}{p_2^2-m_{D}^2}\frac{1}{q^2-m_{D}^2}\mathcal{F}^2(q^2,m_D^2) \nonumber \label{ami} \\ 
&\times&\frac{g_{DDf_0}g_{f_0KK}}{(p_3+p_4)^2-m_{f_0}^2+im_{f_0}\Gamma_{f_0}},  
\end{eqnarray}
\begin{eqnarray}
\mathcal{A}_{(b)}^{D\bar{D}^*}&=&-(i)^3\int\frac{d^4q}{(2\pi)^4}(-\sin\theta g_{\psi_s DD^*}-\cos\theta g_{\psi_d DD^*})g_{J/\psi DD^*} \nonumber \\
&\times&\varepsilon_{\alpha\beta\rho\lambda}(p_5^{\alpha}-p_3^{\alpha}-p_4^{\alpha}-2q^{\alpha})(p_5^{\beta}+p_3^{\beta}+p_4^{\beta})\epsilon_{\psi(4500)}^{\lambda} \nonumber \\
&\times& \varepsilon_{\tau\omega\kappa\mu} (p_{5\kappa}-2q_{\kappa})p_5^{\tau}\epsilon_{J/\psi}^{\omega*}\frac{1}{p_1^2-m_{D}^2}\frac{\tilde{g}^{\rho\mu}(p_2)}{p_2^2-m_{D^*}^2}\frac{1}{q^2-m_{D}^2} \nonumber \\
&\times&\frac{g_{DDf_0}g_{f_0KK}}{(p_3+p_4)^2-m_{f_0}^2+im_{f_0}\Gamma_{f_0}} \mathcal{F}^2(q^2,m_D^2),  
\end{eqnarray}
\begin{eqnarray}
\mathcal{A}_{(c)}^{D^*\bar{D}}&=&(i)^3\int\frac{d^4q}{(2\pi)^4}(-\sin\theta g_{\psi_s DD^*}-\cos\theta g_{\psi_d DD^*})g_{J/\psi DD^*} \nonumber \\
&\times&\varepsilon_{\alpha\beta\rho\lambda}(p_5^{\alpha}-p_3^{\alpha}-p_4^{\alpha}-2q^{\alpha})(p_5^{\beta}+p_3^{\beta}+p_4^{\beta})\epsilon_{\psi(4500)}^{\lambda} \nonumber \\
&\times& \varepsilon_{\tau\omega\kappa\mu} (p_{5\kappa}-2q_{\kappa})p_5^{\tau}\epsilon_{J/\psi}^{\omega*}\frac{\tilde{g}^{\rho\xi}(p_1)}{p_1^2-m_{D^*}^2}\frac{1}{p_2^2-m_{D}^2}\frac{\tilde{g}^{\mu}_{\xi}(q)}{q^2-m_{D^*}^2} \nonumber \\
&\times&\frac{g_{D^*D^*f_0}g_{f_0KK}}{(p_3+p_4)^2-m_{f_0}^2+im_{f_0}\Gamma_{f_0}} \mathcal{F}^2(q^2,m_{D^*}^{2}),   
\end{eqnarray}
\begin{eqnarray}
\mathcal{A}_{(d)}^{D^*\bar{D}^*}&=&(i)^3\int\frac{d^4q}{(2\pi)^4}(-\sin\theta g_{\psi_s D^*D^*}+\cos\theta g_{\psi_d D^*D^*})g_{J/\psi D^*D^*}  \nonumber \\
&\times&[\mathcal{S}(p_{5\mu}-p_{3\mu}-p_{4\mu}-2q_{\mu})g_{\gamma\nu}+(p_{3\gamma}+p_{4\gamma}+q_{\gamma})g_{\mu\nu} \nonumber \\
&-& (p_{5\nu}-q_{\nu})g_{\mu\gamma}]\epsilon_{\psi(4500)}^{\mu}[(p_{5\rho}-2q_{\rho})g_{\omega\alpha}+q_{\omega}g_{\rho\alpha} \nonumber \\
&-&(p_{5\alpha}-q_{\alpha})g_{\rho\omega}]\epsilon_{J/\psi}^{\rho*}\frac{\tilde{g}^{\nu\lambda}(p_1)}{p_1^2-m_{D^*}^2}\frac{\tilde{g}^{\gamma\omega}(p_2)}{p_2^2-m_{D^*}^2}\frac{\tilde{g}^{\alpha}_{\lambda}(q)}{q^2-m_{D^*}^2} \nonumber \\
&\times&\frac{g_{D^*D^*f_0}g_{f_0KK}}{(p_3+p_4)^2-m_{f_0}^2+im_{f_0}\Gamma_{f_0}} \mathcal{F}^2(q^2,m_{D^*}^{2}),  \label{amf} 
\end{eqnarray}
where $p_1=p_3+p_4+q$ and $p_2=p_5-q$,  and $\tilde{g}^{\mu\nu}(q)=-g^{\mu\nu}+q^{\mu}q^{\nu}/q^2$. For convenience, we define $g_{ DD}=(\sin\theta g_{\psi_s DD}+\cos\theta g_{\psi_d DD})$, $g_{ DD^*}=(-\sin\theta g_{\psi_s DD^*}-\cos\theta g_{\psi_d DD^*})$, $g_{ D^*D^*}=(-\sin\theta g_{\psi_s D^*D^*}+\cos\theta g_{\psi_d D^*D^*})$ and the coefficient $\mathcal{S}=(-\sin\theta g_{\psi_s D^*D^*}+4\cos\theta g_{\psi_d D^*D^*})/(-\sin\theta g_{\psi_s D^*D^*}+\cos\theta g_{\psi_d D^*D^*})$. The mass  and width  of the scalar meson $f_0(980)$ are taken to be 993 MeV and 60 MeV \cite{ParticleDataGroup:2022pth}, respectively.  In addition, we introduce a monopole form factor $\mathcal{F}(q^2,m_{D^{(*)}}^{2})$ to describe the off shell effect of the exchanged charmed meson and to regularize the divergence of the loop integral \cite{Locher:1993cc,Li:1996yn,Cheng:2004ru}, which has the expression of 
\begin{eqnarray}
\mathcal{F}(q^2,m_{D^{(*)}}^{2})=\frac{m_{D^{(*)}}^{2}-\Lambda^2}{q^2-\Lambda^2},
\end{eqnarray}
where the cutoff $\Lambda$ can be parameterized as $\Lambda=m_{D^{(*)}}+\alpha\Lambda_{\mathrm{QCD}}$ with scale $\Lambda_{\mathrm{QCD}}=0.22$ GeV.  The denominator $(q^2-\Lambda^2)$ in the monopole form factor can suppress the contribution from large loop momentum, similar to the Pauli-Villas regularization scheme, which ensures the convergence of the loop integral.  Here, the dimensionless parameter $\alpha$ is usually of order one \cite{Cheng:2004ru}.

\begin{figure}[t]
	\includegraphics[width=7.5cm,keepaspectratio]{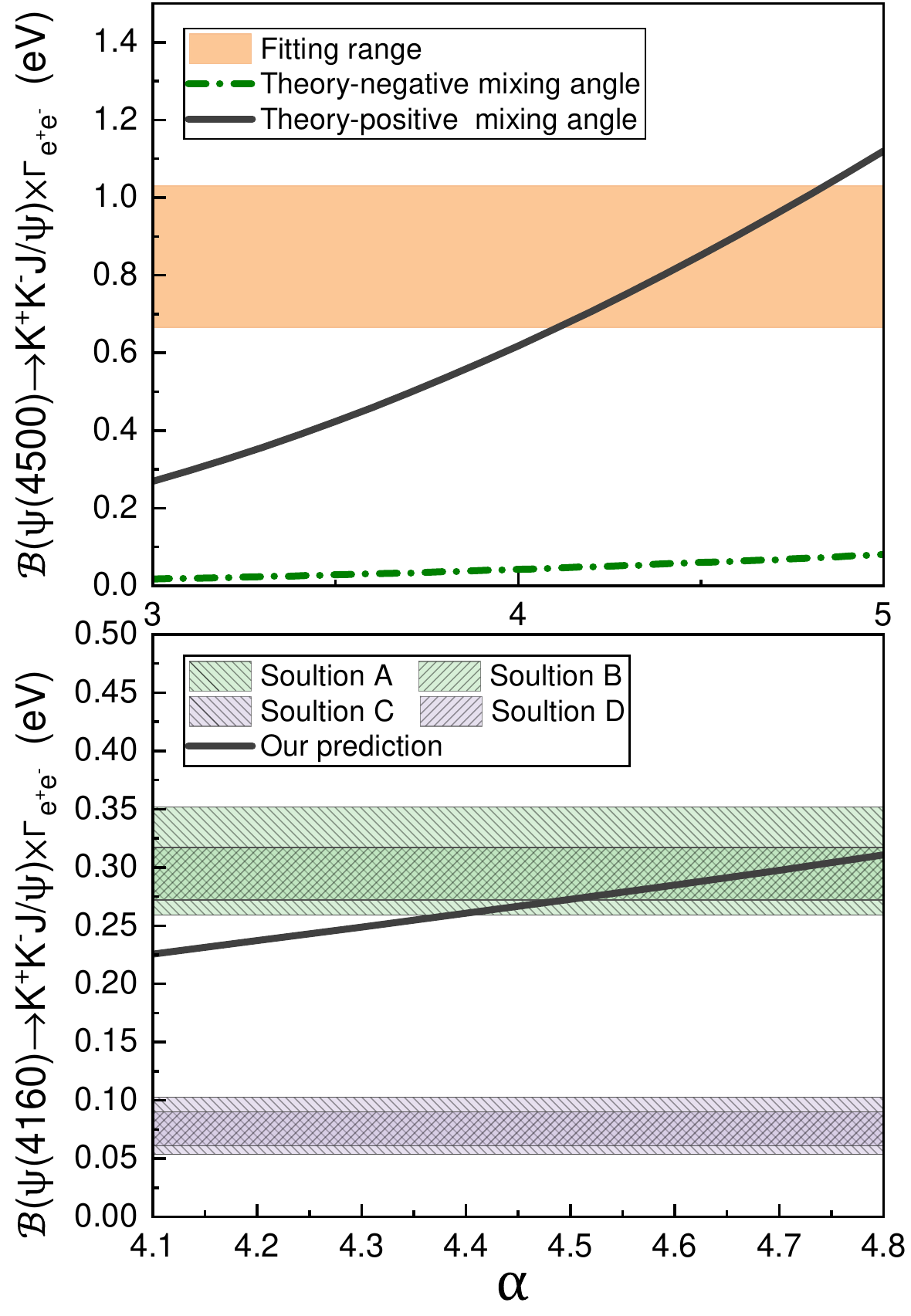}
	\caption{ The $\alpha$ dependence of the product of the di-lepton width $\Gamma(\psi \to e^+e^-)$ and the predicted branching ratio $\mathcal{B}(\psi \to K^+K^- J/\psi)$, where $\psi=\psi(4500)$ and $ \psi(4160)$ have a dominant $D$-wave component. The bands refer to the corresponding fitted solutions of $\Gamma(\psi \to e^+e^-) \mathcal{B}(\psi \to K^+K^- J/\psi)$. }\label{Com4500}
\end{figure}

\begin{table}[b]
  \begin{ruledtabular}
  \caption{ The coupling constants of the interaction between higher charmonia and  charmed meson pairs and di-electron width $\Gamma_{e^+e^-}$ of higher charmonia. }\label{coupling}
  \centering
  \begin{tabular}{ccccccccccc}
     States & $\theta$ & $\Gamma_{e^+e^-}$ (keV)  & $g_{ DD}$ & $g_{DD^{*}}$ (GeV)$^{-1}$ &  $g_{ D^*D^{*}}$ & $\mathcal{S}$ \\
    \hline
    \multirow{2}{*}{$\psi(4500)$} & $+30^{\circ}$ & $0.50$\footnote{They are obtained by combining the theoretical ratio of $\Gamma(\psi(4500)\to e^+e^-)/\Gamma(\psi(4415)\to e^+e^-)=0.86 (0.086)$ \cite{Wang:2019mhs} for the  positive(negative) mixing angle and experimental central value of $\Gamma(\psi(4415)\to e^+e^-)=0.58$ keV \cite{ParticleDataGroup:2022pth}. } & $0.44$ & $-0.076$ &-0.015&-35\\
    & $-30^{\circ}$ & $0.05^a$ & $1.00$ & $-0.013$ & 0.34 &1.70\\
     \cline{1-7}
     $\psi(4415)$ & $+30^{\circ}$ & $0.58\pm0.07$ \cite{ParticleDataGroup:2022pth}  & $0.68$ & $0.003$ &0.43 & 1.71 \\
    $\psi(4160)$ & $+20^{\circ}$\footnote{We also introduce an $S$-$D$ mixing scheme for the established charmonium $\psi(4160)$, whose wave function can be expressed as $-\sin\theta|3^3S_1\rangle+\cos\theta|2^3D_1\rangle$, where $\theta=20^{\circ}$ can be fixed by fitting experimental data of di-electron width of the $\psi(4160)$ \cite{BES:2007zwq}. } & $0.48\pm0.22$ \cite{BES:2007zwq} & $0.47$ & $-0.32$ & 1.17 &4.00\\
    $\psi(4220)$ & $+34^{\circ}$ & 0.29 \cite{Wang:2019mhs} & $0.76$ & $0.054$ &1.22 &1.47\\
     $\psi(4380)$ & $+34^{\circ}$ &  0.26 \cite{Wang:2019mhs}& $0.57$ & $-0.15$ &-0.41 & -1.04 \\     
  \end{tabular}
  \end{ruledtabular}
\end{table}

To calculate the branching ratio of $\psi(4500) \to K^+K^- J/\psi$,  we sum the decay amplitudes in Eqs. (\ref{ami})-(\ref{amf}) and the total amplitude can be given by
\begin{eqnarray}
\mathcal{A}_{\mathrm{Total}}=2\mathcal{A}_{(a)}^{D\bar{D}}+4\mathcal{A}_{(b)}^{D\bar{D}^*}+4\mathcal{A}_{(c)}^{D^*\bar{D}}+2\mathcal{A}_{(d)}^{D^*\bar{D}^*},
\end{eqnarray}
and the branching ratio of $\psi(4500) \to K^+K^- J/\psi$ can be estimated by
\begin{eqnarray}
\mathcal{B}=\frac{\int|\bold{p}_3||\bold{p}_4^*||\mathcal{A}_{\mathrm{Total}}|^2dm_{K^+K^-}d\Omega_3d\Omega_4^*
}{(2\pi)^5 16m_{\psi(4500)}^2\Gamma_{\psi(4500)}},
\end{eqnarray}
where $\bold{p}_4^*$ and $\Omega_4^*$ are the corresponding three-momentum and solid angles of the $K^-$ meson in the center-of-mass frame of the $K^+K^-$ system, respectively.

 \subsection{Numerical results and discussion}

In this section, the calculations of the charmed meson loop contributions to the branching ratio of $\psi \to K^+K^- J/\psi$ depend only on the coupling constants $g_{ D^{(*)}D^{(*)}}$ and the cutoff parameter $\alpha$.  The coupling constants $g_{ D^{(*)}D^{(*)}}$ can be determined by the corresponding experimental or theoretical partial widths of $\psi \to D^{(*)}\bar{D}^{(*)}$ \cite{Wang:2019mhs}, which are  summarized in Table \ref{coupling}. It should be noted that the sign of the mixing angle $\theta$ in the $5S$-$4D$ mixture is not well known in Ref. \cite{Wang:2019mhs}, so we performed the calculations of $\mathcal{B}(\psi(4500) \to K^+K^- J/\psi)$ within positive and negative mixing angle simultaneously. To understand the fitted solutions of $\Gamma(\psi \to e^+e^-) \mathcal{B}(\psi \to K^+K^- J/\psi)$ listed in Table \ref{parameter}, we also need the corresponding di-electron widths $\Gamma(\psi \to e^+e^-)$ of higher charmonia, which are listed in Table \ref{coupling}. Here, experimental measurements on the di-lepton widths of 
two well-established charmonia, $\psi(4160)$ and $\psi(4415)$, were available
 \cite{ParticleDataGroup:2022pth}. And for other charmonium states, we used theoretical estimates from Ref. \cite{Wang:2019mhs}.

\begin{figure}[t]
	\includegraphics[width=7.5cm,keepaspectratio]{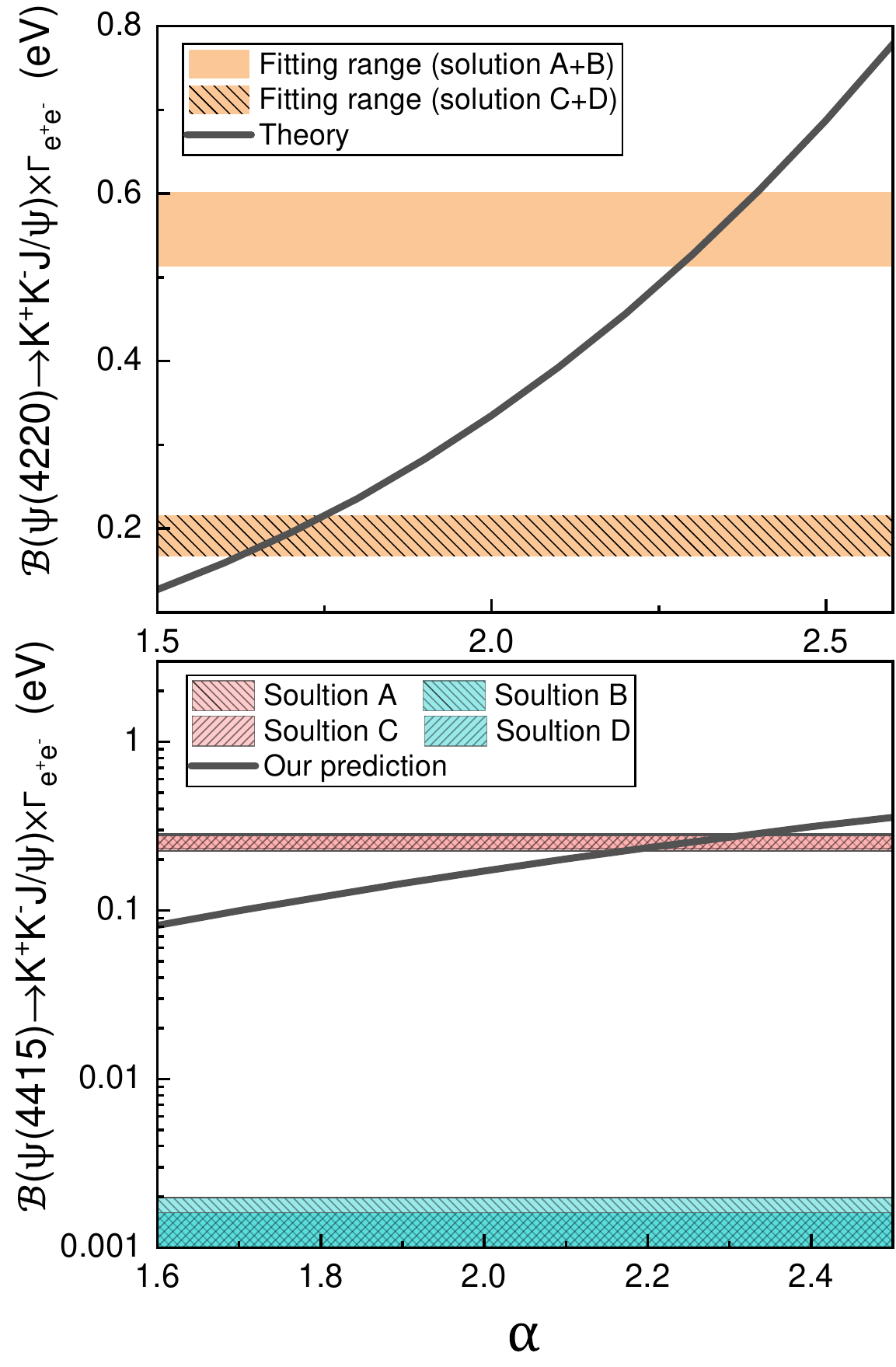}
	\caption{ The $\alpha$ dependence of the product of the di-lepton width $\Gamma(\psi \to e^+e^-)$ and the predicted branching ratio $\mathcal{B}(\psi \to K^+K^- J/\psi)$, where $\psi=\psi(4220)$ and $\psi(4415)$ have a dominant $S$-wave component. The bands refer to the corresponding fitted solutions of $\Gamma(\psi \to e^+e^-) \mathcal{B}(\psi \to K^+K^- J/\psi)$.   }\label{Com4220}
\end{figure}

In Fig. \ref{Com4500}, we show the $\alpha$ dependence of the product of the di-lepton width $\Gamma(\psi(4500) \to e^+e^-)$ and the calculated branching ratio $\mathcal{B}(\psi(4500) \to K^+K^- J/\psi)$ in the case of positive and negative mixing angles. It can be seen that the fitted  $\Gamma(\psi(4500) \to e^+e^-) \mathcal{B}(\psi(4500) \to K^+K^- J/\psi)$ can be explained by the theoretical results of the positive mixing angle in a reasonable cutoff range of $\alpha=4.1\sim4.8$, where four fitted solutions are close to each other without distinguishing.  This indicates that the decay calculations further support the possibility of the new structure around 4.5 GeV in $e^+e^- \to K^+K^- J/\psi$ as the predicted charmonium state $\psi(4500)$, where the positive mixing angle is more favorable.  Furthermore, due to the same decay mechanism, the obtained cutoff $\alpha=4.1\sim4.8$ can be treated as a good scaling range  to predict the hidden charm decays of $\psi \to K^+K^- J/\psi$ for  other higher charmonium states, which gives the charmed meson loop mechanism some predictive power.  However, it should be emphasized that  the cutoff region associated with an $S$-wave dominated charmonium may be very different from that of a $D$-wave dominated charmonium in the charmed meson loop mechanism as shown in Ref. \cite{Qian:2021gby}, so we used the cutoff range of $\alpha=4.1\sim4.8$ to predict the $\Gamma(\psi(4160) \to e^+e^-)\mathcal{B}(\psi(4160) \to K^+K^- J/\psi)$, which was calculated to be $0.226\sim0.311$ eV as shown in Fig. \ref{Com4500}. It can be seen that this theoretical value agrees with the corresponding fitted value of $0.306\pm0.047$ eV in solution $A$ and $0.295\pm0.022$ eV in solution $B$.

We further investigated the charmed meson loop contributions to the branching ratio $\mathcal{B}(\psi(4220) \to K^+K^- J/\psi)$  and found that the calculated product of $\Gamma(\psi(4220) \to e^+e^-)$ and $\mathcal{B}(\psi(4220) \to K^+K^- J/\psi)$  is greater than any of the four solutions when adopting $\alpha=4.1\sim4.8$. Considering the situation of the $\psi(4220)$ as an $4S$-wave dominated charmonium, we need to choose a new scaling region. In Fig. \ref{Com4220}, we show the $\alpha$ dependence of the calculated $\Gamma(\psi(4220) \to e^+e^-)\mathcal{B}(\psi(4220) \to K^+K^- J/\psi)$, where the bands of representing the four fitted solutions may coincide with our theoretical curves within $\alpha=2.3\sim2.4$ or $\alpha=1.6\sim1.7$. Thus, the experimental information in $K^+K^- J/\psi$ channel also agrees to assign the $Y(4220)$ structure to the $J/\psi$ family as a higher charmonium candidate with a dominant $4S$-wave component. Based on the above cutoff scaling region between $\alpha=1.6-2.4$, we  predicted the $\Gamma(\psi(4415) \to e^+e^-)\mathcal{B}(\psi(4415) \to K^+K^- J/\psi)=0.082-0.312$ eV as presented in Fig. \ref{Com4220}, where the $\psi(4415)$ is also an $S$-wave dominated state. It can be seen that this predicted range can match the corresponding fitted values of $0.259\pm0.026$ eV in solution $A$ and $0.252\pm0.027$ eV in solution $C$, and the fitted solutions of $(6.440\pm13.419)\times10^{-4}$ eV in solution $B$ and $(6.343\pm9.804)\times10^{-4}$ eV in solution $D$ can be safely excluded. An interesting phenomenon is that there is no obvious resonance peak of the $\psi(4415)$ in the cross section distribution of $e^+e^- \to K^+K^- J/\psi$ in Fig. \ref{Fit}, and yet our theoretical results favour a comparable solution of $\Gamma(\psi(4415) \to e^+e^-)\mathcal{B}(\psi(4415) \to K^+K^- J/\psi)$ over those of the $\psi(4220)$ and $\psi(4500)$ in $e^+e^- \to K^+K^- J/\psi$, where the interference effect between the resonances changes the line shape of the $\psi(4415)$ signal.

Combining this with the theoretical predictions for the two branching ratios $\mathcal{B}(\psi(4160) \to K^+K^- J/\psi)$ and $\mathcal{B}(\psi(4415) \to K^+K^- J/\psi)$, we can conclude that only solution $A$ in the  fitted products of di-lepton width $\Gamma(\psi \to e^+e^-)$ and branching ratio $\mathcal{B}(\psi \to K^+K^- J/\psi)$ is favoured by theoretical calculations. Therefore, several reliable branching ratios can be obtained, i.e.,
\begin{eqnarray}
\mathcal{B}(\psi(4160) \to K^+K^- J/\psi)=(6.38\pm0.98)\times10^{-4}, \label{bri}
\end{eqnarray}
\begin{eqnarray}
\mathcal{B}(\psi(4220) \to K^+K^- J/\psi)=(1.92\pm0.16)\times10^{-3},  
\end{eqnarray}
\begin{eqnarray}
\mathcal{B}(\psi(4415) \to K^+K^- J/\psi)=(4.47\pm0.45)\times10^{-4},
\end{eqnarray}
\begin{eqnarray}
\mathcal{B}(\psi(4500) \to K^+K^- J/\psi)=(1.83\pm0.23)\times10^{-3},  \label{brf}
\end{eqnarray} 
which should provide valuable information for guiding experiments to search for these vector charmonium states in other production platforms, such as $B$ meson decay, proton-proton collisions, etc.

\subsection{Possibility of finding out another predicted charmonium $\psi(4380)$ in $e^+e^- \to   K^+K^- J/\psi$ }

\begin{figure}[b]
	\includegraphics[width=8.0cm,keepaspectratio]{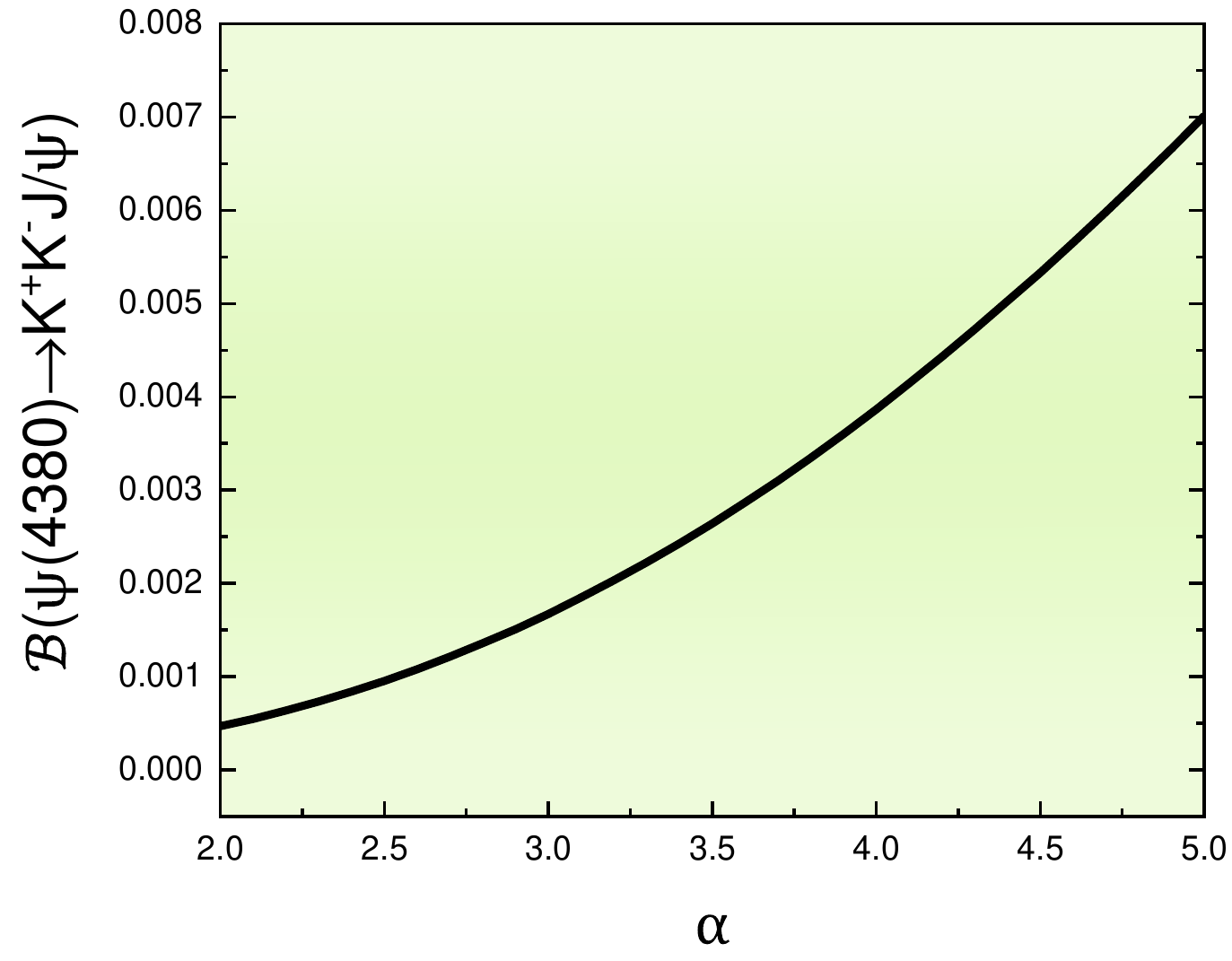}
	\caption{ The predictions for the branching ratio $\mathcal{B}(\psi(4380) \to K^+K^- J/\psi)$.   }\label{Br4380}
\end{figure}

As mentioned in the Introduction, in addition to the $\psi(4500)$ state, it should be noticed that  there is another charmonium $\psi(4380)$ predicted as a partner state of the $\psi(4220)$ in our spectroscopy framework \cite{Wang:2019mhs}. The possibility of discovering this charmonium state in $e^+e^- \to   K^+K^- J/\psi$ should then be investigated.  In this subsection, we discuss the resonance contribution of the charmonium $\psi(4380)$ in $e^+e^- \to K^+K^- J/\psi$. The relevant information of the $\psi(4380)$ is also summarized in Tables \ref{masswidth} and \ref{coupling}.

\begin{figure}[t]
	\includegraphics[width=8.0cm,keepaspectratio]{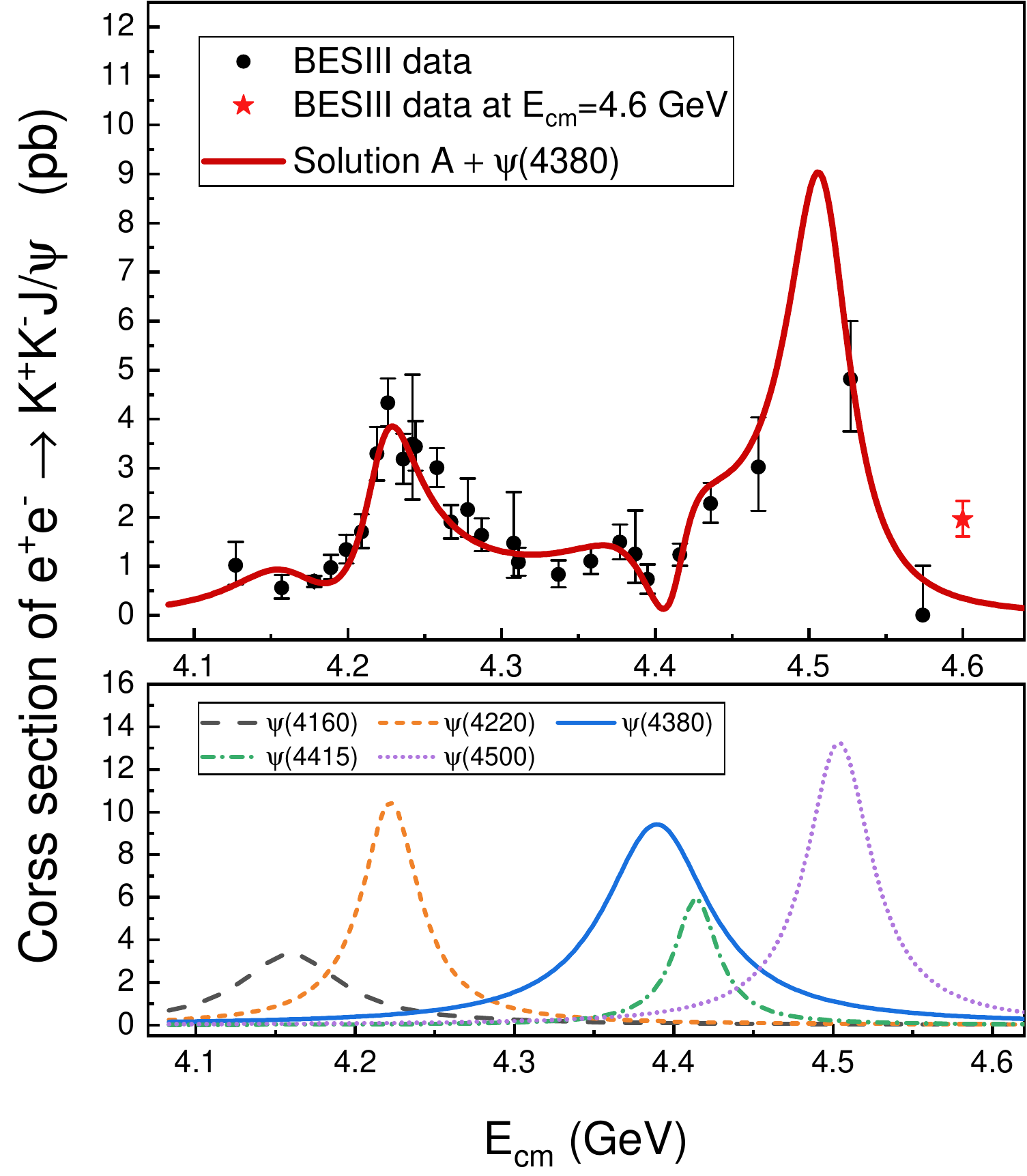}
	\caption{The fit to the cross section of $e^+e^- \to   K^+K^- J/\psi$ based on the solution $A$ plus the predicted resonance contribution of charmonium $\psi(4380)$.   }\label{fit4380}
\end{figure}

The predictions for the branching ratio $\mathcal{B}(\psi(4380) \to K^+K^- J/\psi)$ are shown in Fig. \ref{Br4380}, which can reach up to the order of magnitude of  $10^{-3}$. Considering the $D$-wave dominance in the wave function of the $\psi(4380)$, we can obtain the branching ratio 
\begin{eqnarray}
\mathcal{B}(\psi(4380) \to K^+K^- J/\psi)=(4.0\sim6.0)\times10^{-3}
\end{eqnarray}
 within $\alpha=4.1\sim4.8$, which is certainly comparable to those in Eqs. (\ref{bri})-(\ref{brf}). Such a large branching ratio means that it is very hopeful to detect the $\psi(4380)$ in the $e^+e^- \to   K^+K^- J/\psi$ process. However, it is confusing that there is no clear peak structure to  support the existence of the $\psi(4380)$ in the present data of the $e^+e^- \to   K^+K^- J/\psi$ cross section as seen in Fig. \ref{Fit}. To understand this point, we have performed a complete fit to the cross section of $e^+e^- \to   K^+K^- J/\psi$ by combining the contribution of solution $A$ and the predicted resonance contribution of charmonium $\psi(4380)$, where the typical value of $\Gamma(\psi(4380) \to e^+e^-)\mathcal{B}(\psi(4380) \to K^+K^- J/\psi)=(260~\mathrm{eV})\times(4\times10^{-3}$)=1.04 eV was used as the input.
 The corresponding fitting results and phase parameters are shown in Fig. \ref{fit4380} and Table \ref{parameter2}, respectively, where it can be seen that the BESIII data  can also be well described. Compared with the fitted line shape of the solution $A$ shown in Fig. \ref{Fit}, the most obvious difference in Fig. \ref{fit4380} is that  there appear an enhancement structure around 4.365 GeV and a dip structure around 4.405 GeV, which just embody the contribution of the $\psi(4380)$. Due to the large bin size of the present data points near 4.4 GeV \cite{BESIII:2022joj}, our predicted signal of the $\psi(4380)$ in $e^+e^- \to   K^+K^- J/\psi$ cannot be directly confirmed, which can be tested by more precise experimental measurements from BESIII or Belle II in the future.

\begin{table}[t]
  \begin{ruledtabular}
  \caption{ The fitted parameters in the scheme of considering the contributions of the solution A and the charmonium $\psi(4380)$. The phase angle $\phi_4$ is relevant to the $\psi(4380)$.  }\label{parameter2}
  \centering
  \begin{tabular}{ccc}
     ~~~Parameters & Values & Error$(\pm)$~~~   \\
    \hline
    $\phi_{1}$ ~(rad) &  $2.378$&$0.066$ ~~~  \\
    $\phi_{2}$ ~(rad) &  $3.328$  & 0.036~~~     \\
    $\phi_{3}$ ~(rad) &  $1.876$ & 0.038   ~~~   \\
    $\phi_{4}$ ~(rad) &  $5.095$ & 0.028    ~~~  \\
    $m_{\psi(4500)}$~~(MeV)  &  $4509$ & 4 ~~~\\
    $m_{\psi(4380)}$~~(MeV)   & $4389$ & 1 ~~  \\
    \hline
    $\chi^2/\mathrm{d.o.f.}$ &  $1.06$      \\
  \end{tabular}
  \end{ruledtabular}
\end{table}

\section{Summary}\label{sec4}

Recently, the BESIII Collaboration had measured the Born cross section of $e^+e^- \to   K^+K^- J/\psi$ from 4.127 to 4.600 GeV, in which two charmoniumlike structures were observed  \cite{BESIII:2022joj}. The first resonance can be related to the famous $Y(4220)$, and the second resonance around 4.5 GeV, denoted by the $Y(4500)$, was observed for the first time.  Surprisingly, the measured mass of  the newly observed $Y(4500)$ can be well matched with that of a candidate charmonium candidate $\psi(4500)$ predicted in our previous work \cite{Wang:2019mhs}. In this spectroscopy framework \cite{Wang:2019mhs}, the $Y(4220)$ and $\psi(4415)$ can be very well assigned to the vector charmonium family under a $4S$-$3D$ and $5S$-$4D$ mixing scheme, respectively. As the predictions, two corresponding partner states $\psi(4380)$ and $\psi(4500)$ were obtained \cite{Wang:2019mhs}.  Therefore, the recent observation of the $Y(4500)$ gives us confidence in identifying the existence of the predicted charmonium $\psi(4500)$.  

To further confirm the nature of the $Y(4500)$, we first reanalyzed the cross section of $e^+e^- \to   K^+K^- J/\psi$ by introducing two theoretically predicted charmonium states $\psi(4220)$ and $\psi(4500)$ as intermediate resonances in the scheme I.  We found that the broader enhancement structure around 4.5 GeV can be reproduced by a narrower $\psi(4500)$ state when the enhanced data point at 4.600 GeV is not included in the fit, which may be caused by the contributions from higher charmonium states above 4.6 GeV  \cite{Wang:2020prx,Qian:2021gby}.
After resolving this width discrepancy problem, we found that the refined description for the cross section of $e^+e^- \to   K^+K^- J/\psi$ in the fit scheme I is still not good, so a four-resonance fit scheme II was introduced, considering the contributions of two well-established charmonium states $\psi(4160)$ and $\psi(4415)$.  We found that the inclusion  of the $\psi(4160)$ and $\psi(4415)$ can obviously improve the fitting quality, with four sets of very different solutions to $\Gamma_{e^+e^-} \mathcal{B}(\psi \to K^+K^- J/\psi)$ were obtained. 

Subsequently, we focused mainly on the charmonium hadronic transition to the  final states $K^+K^- J/\psi$. Using the charmed meson loop mechanism, we calculated the branching ratios $\mathcal{B}(\psi(4500) \to K^+K^- J/\psi)=(0.13\sim0.20)\%$, $\mathcal{B}(\psi(4220) \to K^+K^- J/\psi)=(0.18\sim0.21)\%$, $\mathcal{B}(\psi(4160) \to K^+K^- J/\psi)=(0.047\sim0.065)\%$ and $\mathcal{B}(\psi(4415) \to K^+K^- J/\psi)=(0.047\sim0.054)\%$. Combined with the di-lepton widths of these charmonium states, we found only one suitable solution for the product of $\Gamma(\psi \to e^+e^-)$ and the branching ratio of $\mathcal{B}(\psi \to K^+K^- J/\psi)$, supported by the corresponding theoretical calculation, which further reinforces the possibility of the newly observed structure around 4.5 GeV in $e^+e^- \to   K^+K^- J/\psi$ as a higher charmonium $\psi(4500)$.   Finally, we also investigated the possible signal of another predicted candidate charmonium $\psi(4380)$ in $e^+e^- \to   K^+K^- J/\psi$. We found that the $\psi(4380)$ should play a non-negligible role in the $e^+e^- \to   K^+K^- J/\psi$ process, where the predicted branching ratio $\mathcal{B}(\psi(4380) \to K^+K^- J/\psi)=(4.0\sim6.0)\times10^{-3}$  is larger than those of other charmonium states.  As a testable prediction, we found that the contribution of $\psi(4380)$ can produce an enhancement structure around 4.365 GeV and a dip structure around 4.405 GeV in the cross section distribution of $e^+e^- \to   K^+K^- J/\psi$, which can be checked by the experiments.  Therefore, we suggest our experimental colleague to perform a more precise measurement of the cross section of $e^+e^- \to   K^+K^- J/\psi$, which is crucial to construct the higher charmonium spectrum and to test our proposal for the charmoniumlike $Y$ problem.

Our research results presented here have confirmed the existence of the predicted charmonium $\psi(4500)$ state, which should be an important step towards the construction of the charmonium family. Further experimental and theoretical efforts on the issue of the vector charmoniumlike $Y$ states are still needed in the future.

\section*{ACKNOWLEDGEMENTS}

This work is supported by the China National Funds for Distinguished Young Scientists under Grant No. 11825503, the National Key Research and Development Program of China under Contract No. 2020YFA0406400, the 111 Project under Grant No. B20063, the National Natural Science Foundation of China under Grant No. 12247101, and the Project for top-notch innovative talents of Gansu province. J.-Z.W. is also supported by the National Postdoctoral Program for Innovative Talent.

\end{document}